\documentclass[12pt,a4paper]{article}

\usepackage{graphicx}

\begin{document}
	
\title{SIMULATION OF GAUSSIAN RANDOM FIELD IN A BALL}
	
\author{
D. Kolyukhin \\ 
Trofimuk Institute of Petroleum Geology and Geophysics\\ SB RAS,
pr. Koptyuga, 3, 630090, Novosibirsk, Russia.\\ E-mail: KolyukhinDR@ipgg.sbras.ru\\
\and
A. Minakov\\
Centre for Earth Evolution and Dynamics (CEED),\\ University of Oslo, Sem Sælands vei 2A, 0371 Oslo, Norway. \\ E-mail: alexamin@uio.no
}

%	\date{January 2016}
	
\maketitle
	
\begin{abstract}
		
The presented paper is devoted to statistical modeling of Gaussian scalar real random fields inside a three-dimensional sphere (ball). We propose a statistical model describing the spatial heterogeneity in a unit ball and a numerical procedure for generating an ensemble of corresponding random realizations. The accuracy of the presented approach is corroborated by the numerical comparison of the estimated and analytical covariance functions. Our approach is flexible with respect to the assumed radial and angular covariance function. We illustrate the effect of the covariance model parameters based on numerical examples of random field realizations. The presented statistical simulation technique can be applied, for example, to the inference of the 3D spatial heterogeneity in the Earth and other planets.
		
\end{abstract}


\section{Introduction}

The paper is devoted to modeling spatially inhomogeneous objects with the shape of a ball. For example, this work can be used in planetary sciences or studies focusing on the 3D thermochemical structure of the Earth's mantle. The geophysical and geochemical data suggest that the Earth’s mantle contains heterogeneities on a broad range of scales: from thousands of km down to the grain size \cite{Stixrude:2012}. The stochastic mantle heterogeneity can be attributed to thermal and chemical variations due to mantle convection and melting processes \cite{Allegre:1986}, \cite{Armienti:2010}. As a rule, the complex nature of physical processes, and the lack of measurement data, a fully deterministic description of the problem is impossible. Therefore, efficient statistical modeling methods are required for a better understanding of the formation and preservation of the 3D planetary heterogeneity. This approach involves generating a statistical ensemble of realizations instead of a single deterministic solution. This allows, for example, to estimate the confidence intervals of the desired characteristics, as well as perform uncertainty quantification and sensitivity analyzes (\cite{Urrego-Blanco:2016}; \cite{McClarren:2018}).

Over the past decade, a wide range of mathematical techniques based on the Schoenberg representation of the covariance function has been developed for modeling homogeneous isotropic random fields on a spherical surface (e.g. \cite{Creasey:2018}; \cite{Emery:2019}). On the other hand, volumetric modeling in the full-sphere is not well covered. To our knowledge in the scientific literature, this subject is restricted by the isotropic homogeneous case (\cite{Yaglom:1962}; \cite{Marinucci:2011}). Meschede \& Romanowicz \cite{Meschede:2015} extended the random field simulation technique for the non-homogeneous distribution along the radius using the Karhunen–Loève expansion. However, in our opinion, the disadvantage of this approach is the difficulty in a rigorous mathematical description of the simulated random field. In the presented work, we develop an alternative method for modeling a random field in a ball.

\section{Statistical model}

\subsection{Modeling Gaussian random fields in Cartesian coordinates}

Gaussian random fields are fully determined by mean values and covariance function which in the real case has form
\begin{equation}
\label{eqn:1}
C(\mbox{\boldmath{$x$}},\mbox{\boldmath{$y$}})=E \left \{(f(\mbox{\boldmath{$x$}})-E{f(\mbox{\boldmath{$x$}})})(f(\mbox{\boldmath{$y$}})-E{f(\mbox{\boldmath{$y$}})}) \right \}, \quad \mbox{\boldmath{$x$}}, \mbox{\boldmath{$y$}} \in R^3~.
\end{equation}
Here and below in this work, we use bold font for vectors and matrices, the overbar denotes the complex conjugate. $E$ denotes the mathematical expectation, $f(\mbox{\boldmath{$x$}})$ and $f(\mbox{\boldmath{$y$}})$ are values of random field realization at the coordinates defined by vectors $\mbox{\boldmath{$x$}}$ and $\mbox{\boldmath{$y$}}$ respectively. Further in this work, we deal with zero mean random fields.

In the case when the values of the random field $f(\mbox{\boldmath{$x$}})$ should be generated in the finite number of fixed spatial points, the traditional approach to simulate the realizations is based on the Cholesky decomposition of the covariance matrix $\mbox{\boldmath{$C$}}$
\begin{equation}
\label{eqn:2}
\mbox{\boldmath{$C=LL^T$}}~, \quad \mbox{\boldmath{$f=L\xi$}} ~,
\end{equation}
where $\mbox{\boldmath{$\xi$}}$ is a vector of mutually independent random numbers with zero mean and unit variance, $\mbox{\boldmath{$L^T$}}$ is the transpose of the lower triangular matrix $\mbox{\boldmath{$L$}}$. 

Another approach to statistical modeling uses the Karhunen-Loève expansion
\begin{equation}
\label{eqn:3}	
\mbox{\boldmath{$C = V^T D V $}} ~, \quad \mbox{\boldmath{$f = V^T \sqrt{D} \xi $}}~,
\end{equation}
where $\mbox{\boldmath{$D$}}$ and $\mbox{\boldmath{$V$}}$ are diagonal matrix of eigenvalues and matrix whose columns are the corresponding eigenvectors of the covariance matrix, respectively. In the truncated Karhunen-Loève expansion, only the eigenvectors corresponding to significant eigenvalues are considered.

In practice, the size of the matrix $\mbox{\boldmath{$C$}}$ is often too large, and decomposition in Eqs. \ref{eqn:2},\ref{eqn:3} is computationally consuming due to the need to perform the Cholesky decomposition or to solve the eigenvalues problem in a high dimension. In application to simulation of the 3D random field in a ball, Meschede and Romanowicz \cite{Meschede:2015} suggested to apply the Karhunen-Loève expansion only in the radial direction to avoid this numerical problem. 

\subsection{2D random fields on the sphere}

This subsection is devoted to the modeling of isotropic random fields on the surface of a sphere. Consider the unit sphere $S^2=\{\mbox{\boldmath{$x$}} \in R^3: |\mbox{\boldmath{$x$}}|=1\}$. A basis formed by spherical harmonic functions
\begin{equation}
\label{eqn:4}
Y_n^k (\phi,\theta)=\sqrt{ \frac{2n+1}{4\pi} \frac{(n-k)!}{(n+k)!} } P_n^k (cos(\theta)) e^{ik\phi} 
\end{equation} 
is used to describe such random fields. Here $P_n^k$ are associated Legendre functions, where $n$ and $k$ are the degree and order of spherical harmonics, respectively, such as $n \ge 0$, $-n \le k \le n$ and, $0 \le \phi < 2\pi$ is longitude and, $0 \le \theta \le \pi$ is colatitude. The $S^2$ metric is the geodesic distance which defines the distance between points $\mbox{\boldmath{$x$}}$ and $\mbox{\boldmath{$y$}}$ as angle 
$$
\alpha= arccos \left( \mbox{\boldmath{$x$}} \cdot \mbox{\boldmath{$y$}} \right) ~,
$$
where
$$
\mbox{\boldmath{$x$}}= (cos(\phi_x) sin(\theta_x), sin(\phi_x) sin(\theta_x), cos(\theta_x)) ~,
$$
$$
\mbox{\boldmath{$y$}}= (cos(\phi_y) sin(\theta_y ),sin(\phi_y) sin(\theta_y), cos(\theta_y)) ~.
$$
Then any complex function $f \in L^2 (S^2)$ can be presented as
\begin{equation}
\label{eqn:5}
f(\mbox{\boldmath{$x$}})=f(\phi,\theta)=\sum_{n=0}^\infty \sum_{k=-n}^n f_n^k Y^k(\phi,\theta) ~,  
\end{equation}
where due to the orthonormality of the basis of spherical harmonics coefficients $f_n^k$ has the form

$$
f_n^k=\int_0^{2\pi} \int_0^\pi f(\phi,\theta) \overline{Y_n^k(\phi,\theta)} d\theta d\phi .
$$
The spherical harmonics satisfy the important property for two arbitrary points on the spherical surface defined by vectors $\mbox{\boldmath{$x$}}$ and $\mbox{\boldmath{$y$}}$
$$
\frac{4\pi}{2n+1} \sum_{k=-n}^n Y_n^k(\mbox{\boldmath{$x$}}) \overline{Y_n^k(\mbox{\boldmath{$y$}})} =  P_n(\mbox{\boldmath{$x$}} \cdot \mbox{\boldmath{$y$}}),~\mbox{\boldmath{$x$}},\mbox{\boldmath{$y$}} \in S^2   
$$
where $P_n=P_n^0$ is a Legendre polynomial.

We consider $f(\mbox{\boldmath{$x$}})$ as a real isotropic random field with zero mean on $S^2$.  $f(\mbox{\boldmath{$x$}})$ is isotropic if the covariance at two points $\mbox{\boldmath{$x$}}$ and $\mbox{\boldmath{$y$}}$
$$
Cov(f(\mbox{\boldmath{$x$}}),f(\mbox{\boldmath{$y$}}))=C_\alpha (\alpha(\mbox{\boldmath{$x$}},\mbox{\boldmath{$y$}})), \mbox{\boldmath{$x$}},\mbox{\boldmath{$y$}} \in S^2   
$$
is finite and depends only on the geodesic distance between them. Here $\alpha$ is an angle between points $\mbox{\boldmath{$x$}}$, $\mbox{\boldmath{$y$}}$ on a sphere, i.e. $cos(\alpha)=\mbox{\boldmath{$x$}} \cdot \mbox{\boldmath{$y$}}$.

According to Schoenberg theorem \cite{Schoenberg:1942}, $C_\alpha$ can be a covariance function of an isotropic random field on the sphere if and only if

\begin{equation}
\label{eqn:6}
C_\alpha (\alpha)=\sum_{n=0}^\infty a_n P_n cos(\alpha)~, \alpha \in [0,\pi]
\end{equation}                                     
for some summable series of non-negative coefficients $\{a_n\}$ \cite{Lantuejoul:2019}. A detailed survey of known functions and corresponding decompositions can be found, for example, in \cite{Terdik:2015} and \cite{Lantuejoul:2019}. In practice, the summation from $0$ to a finite spherical harmonic degree number $n_{max}$ is used in Eq. \ref{eqn:6}. The numerical realizations of the random field on the sphere can be obtained using the mathematical apparatus described in \cite{Creasey:2018}, \cite{Emery:2019}, and \cite{Lantuejoul:2019}.

\subsection{3D random fields in the ball}

In this subsection, we consider the random fields distributed in the 3D filled sphere (ball). The modeling approach is based on using the covariance function assumed proportional to the product of the radial and isotropic angular covariance functions
\begin{equation}
\label{eqn:7}
C(\mbox{\boldmath{$x$}},\mbox{\boldmath{$y$}})=E\{f(\mbox{\boldmath{$x$}})f(\mbox{\boldmath{$y$}})\} \sim C_r (r_x,r_y ) C_\alpha(\alpha)~.               
\end{equation}
Here \mbox{\boldmath{$x$}} and \mbox{\boldmath{$y$}} the pair of points in spherical coordinates
$$
\mbox{\boldmath{$x$}}= (r_x cos(\phi_x) sin(\theta_x), r_x sin(\phi_x) sin(\theta_x), r_x cos(\theta_x)) ~,
$$
$$
\mbox{\boldmath{$y$}}= (r_y cos(\phi_y) sin(\theta_y), r_y sin(\phi_y) sin(\theta_y ), r_y cos(\theta_y))  ~.
$$
$C_r$ is a covariance function in Cartesian coordinates and $C_\alpha$ is a covariance function on a sphere that has the form Eq. \ref{eqn:7} and
$$
\alpha= arccos \left( \frac{\mbox{\boldmath{$x$}}}{r_x} \cdot \frac{\mbox{\boldmath{$y$}}}{r_y} \right) ~.
$$
This representation of the 3D covariance function with Eq. \ref{eqn:7} allows a rigorous mathematical description of a radially inhomogeneous volumetric random field in a full sphere. In contrast to the statistical method developed by Meschede \& Romanowicz \cite{Meschede:2015}, the angular distribution for the second method is explicitly described by the covariance function $C_\alpha$. Moreover, both the angular and radial covariance functions can be estimated based on regional or global geophysical observations.

\section{Simulation method}

The random field realizations $f(\mbox{\boldmath{$x$}})$ on a sphere can be generated using the spectral method developed in \cite{Lantuejoul:2019}. Simulation formula has the form

\begin{equation}
\label{eqn:8}
f(\mbox{\boldmath{$x$}})=\frac{1}{\sqrt{N}} \sum_{l=1}^N f_l(\mbox{\boldmath{$x$}}) ~.  
\end{equation}

The central limit theorem ensures that for a large number of realizations $N$ and independent identically distributed $f_l$, Eq. \ref{eqn:8} generates a Gaussian random field.

Each realization of $f_l$ is generated according to the following algorithm:
\begin{itemize}
\item
generate an integer random number according to distribution ${a_n}$ in Eq. \ref{eqn:6}
$$
n_l \sim {a_n};
$$
\item
generate an integer random number uniformly distributed in the interval $[-n_l,n_l]$
$$
k_l \sim U[-n_l,n_l];
$$
\item
the resulting formula for modeling $f_l$ is
$$
f_l(\mbox{\boldmath{$x$}})=2\sqrt{\pi A} \left( \xi Re(Y_{n_l}^{k_l} (\mbox{\boldmath{$x$}})) + \eta Im(Y_{n_l}^{k_l} (\mbox{\boldmath{$x$}})) \right)  ,
$$
\end{itemize}
where $\xi$ and $\eta$ are the mutually independent random numbers with zero mean and unit variance and $A=\sum_0^{n_{max}} a_n$.

For generalization to the three-dimensional case with covariance function defined by Eq. \ref{eqn:7}, we suggest a method based on the Cholesky decomposition of the radial covariance matrix $\mbox{\boldmath{$C_r=L_r L_r^T$}}$ using Eq. \ref{eqn:2} or Karhunen-Loève expansion $\mbox{\boldmath{$C_r=V_r^T D_r  V_r$}}$ using Eq. \ref{eqn:3}. Matrix $\mbox{\boldmath{$C_r$}}$ is defined as the values of function $C_r$ in Eq. \ref{eqn:7} for all pairs of radius values $r_i,i=1,\ldots,M$.

The resulting simulation formula is
\begin{equation}
\label{eqn:9}
f(r_i,\phi,\theta) = \frac{2\sqrt{\pi A}}{\sqrt{N}} \sum_{l=1}^N \left( \xi(r_i ) 
Re(Y_{n_l}^{k_l} (\phi,\theta)) + \eta(r_i ) 
Im(Y_{n_l}^{k_l} (\phi,\theta)) \right) ~, 
\end{equation}

where

\begin{equation}
\label{eqn:10}
\mbox{\boldmath{$\xi=L_r \xi_0$}}~, \mbox{\boldmath{$\eta=L_r \eta_0$}}~,
\end{equation}

or

\begin{equation}
\label{eqn:11}
\mbox{\boldmath{$\xi=V_r^T \sqrt{D_r} \xi_0$}}~, \mbox{\boldmath{$\eta=V_r^T \sqrt{D_r} \eta_0$}}~.                                           
\end{equation}

Here, $\mbox{\boldmath{$\xi_0$}}$, $\mbox{\boldmath{$\eta_0$}}$ are the vectors of mutually independent Gaussian random numbers with zero mean and unit variance. According to the central limit theorem, the summation over sufficiently large $N$ in Eq. \ref{eqn:9} provides the Gaussian distribution of the simulated random field. The algorithm described above ensures the reproducing of the covariance function $C(\mbox{\boldmath{$x$}},\mbox{\boldmath{$y$}})$, which proves the positive definiteness of the product in Eq. \ref{eqn:7}. 

\section{Numerical results}

In this section, we present the results of statistical modeling performed by the method described above. In this paper, we present numerical examples for the homogeneous exponential covariance function
$$
C_r(d)=C_r(r_x,r_y )=\sigma^2 exp(\frac{-|r_x-r_y |}{I})~,
$$
where $\sigma$ is a standard deviation, and $I$ is a correlation length of the random field. The angular covariance $C_\alpha$ has the form \cite{Terdik:2015}
$$
C_\alpha (\alpha)=\sum_{n=0}^\infty \rho^n P_n (cos(\alpha)) = 
\frac{1}{\sqrt{1-2 \rho cos(\alpha)+\rho^2}}, \quad
a_n=\rho^n, \rho \in (0,1)~. 
$$
Here the coefficients $a_n$ define the spectrum $\{a_n\}$ in Eq. \ref{eqn:6}.

In Fig.1 we compare the covariance function $C$ calculated on the segment between points $(r_x=0.5,\phi_x=\pi/6,\theta_x=\pi/6)$ and  $(r_y=1,\phi_y=\pi/2,\theta_y=\pi/2)$
with Eqs. \ref{eqn:9}-\ref{eqn:11} and the analytical covariance function Eq. \ref{eqn:7}. The covariance function estimated based on spatial realizations coincides with the analytical curve. It confirmes that the numerical realizations accurately reproduce the assumed statistical model. These results confirm the numerical accuracy of the method described above.

\begin{figure}
	\begin{center}	
		\includegraphics[width=.9\textwidth]{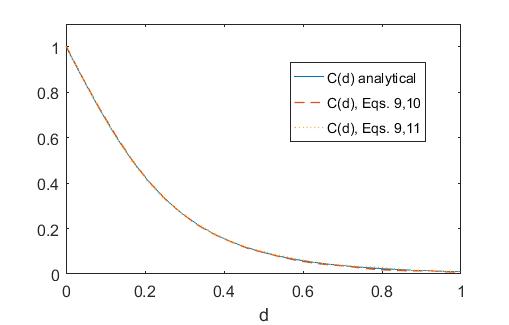}
		\caption{The covariance function $C$ obtained using Eq. (9) and analytical covariance function (7) for $I=0.15$ and $\rho=0.7$).}\label{fig:fig1}
	\end{center}
\end{figure}

In Fig. 2, we show the realizations of a random field using a 2D cross-section (left plot)  and spherical surface (right plot)  assuming $I=0.05$ and $\rho=0.9$. In Fig.3, we present the 2D cross-sections for different values $I=0.25$ and $\rho=0.9$ (left plot) and $I=0.05$ and $\rho=0.6$ (right plot). Comparison of Figs. 2 and 3 demonstrates the radial influence of parameter $I$ (Fig. 3, left plot) and the angular influence of parameter $\rho$ (Fig. 3, right plot). In particular, Fig. 3 shows that the tuning of the parameters of the second statistical model also allows for modeling anisotropic random fields. 

In Fig. 4, we show the 3D realizations of a random field for $I=0.05$ and $\rho=0.7$ using orthographic projection. Eqs. \ref{eqn:9} and \ref{eqn:11} were used to generate the random field realization in Fig. 4. The summation in Eq. \ref{eqn:11} was carried out for the eigenvectors $\mbox{\boldmath{$v_i$}}$ corresponding to $M_1$ largest eigenvalues $\lambda_i$ consisting up 95\% of the trace of the covariance matrix $\mbox{\boldmath{$C_r$}}$
$$
\mbox{\boldmath{$\xi$}}=\sum_{i=1}^{M_1} \sqrt{\lambda_i} \mbox{\boldmath{$v_i$}} \xi_i ~,
$$
where $\{\mbox{\boldmath{$v_i$}}\}$ is the eigenvectors corresponding $\{\lambda_i\}$ and $\{\xi_i\}$ is an independent standard Gaussian random values.

\begin{figure}
	\begin{center}	
		\includegraphics[width=.475\textwidth]{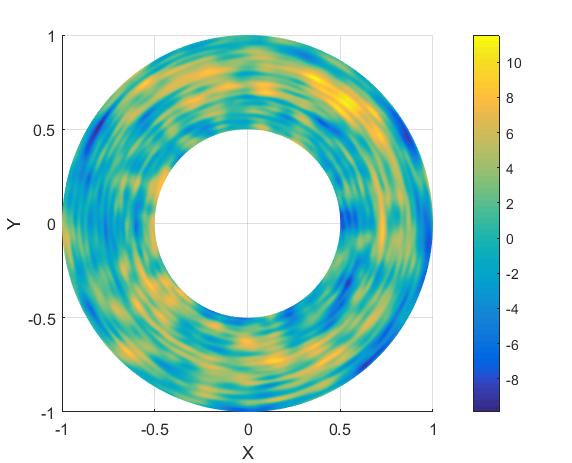}
		\includegraphics[width=.475\textwidth]{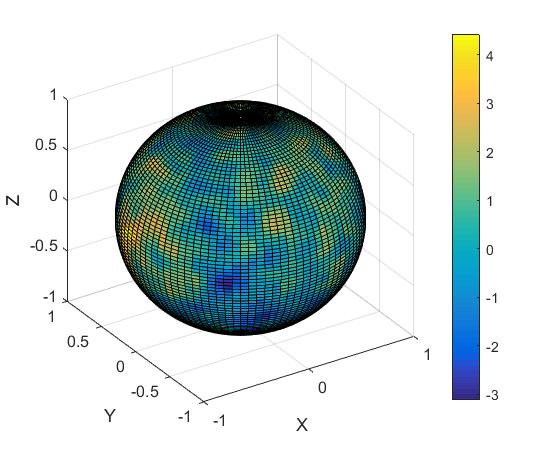}		
		\caption{2D cross-section (left) and spherical surface (right) of 3D realizations of random field related to the statistical model for $I=0.05$ and $\rho=0.9$.}\label{fig:fig2}
	\end{center}
\end{figure}

\begin{figure}
	\begin{center}	
		\includegraphics[width=.475\textwidth]{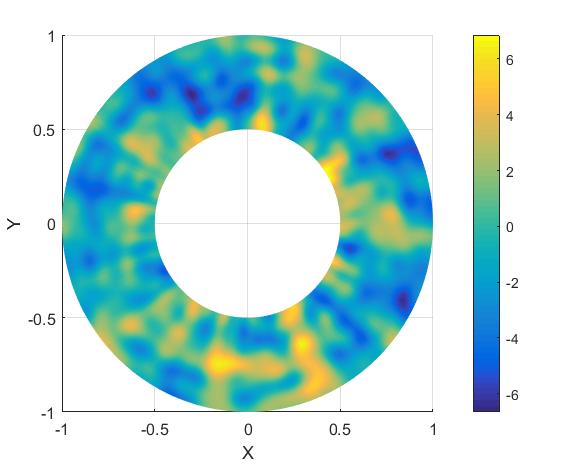}
		\includegraphics[width=.475\textwidth]{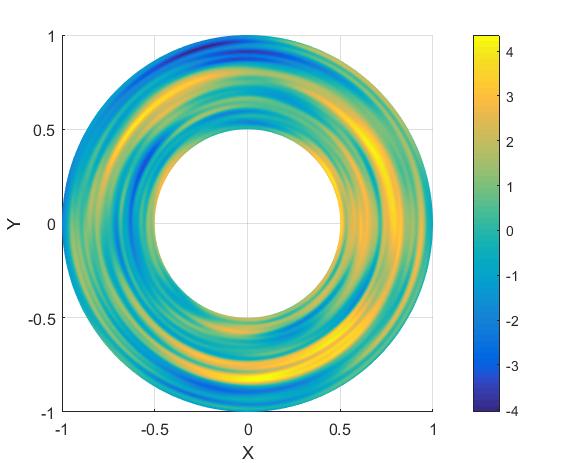}		
		\caption{2D cross-sections for $I=0.25$ and $\rho=0.9$ (left plot) and $I=0.05$ and $\rho=0.6$ (right plot).}\label{fig:fig3}
	\end{center}
\end{figure}

\begin{figure}
	\begin{center}	
		\includegraphics[width=.95\textwidth]{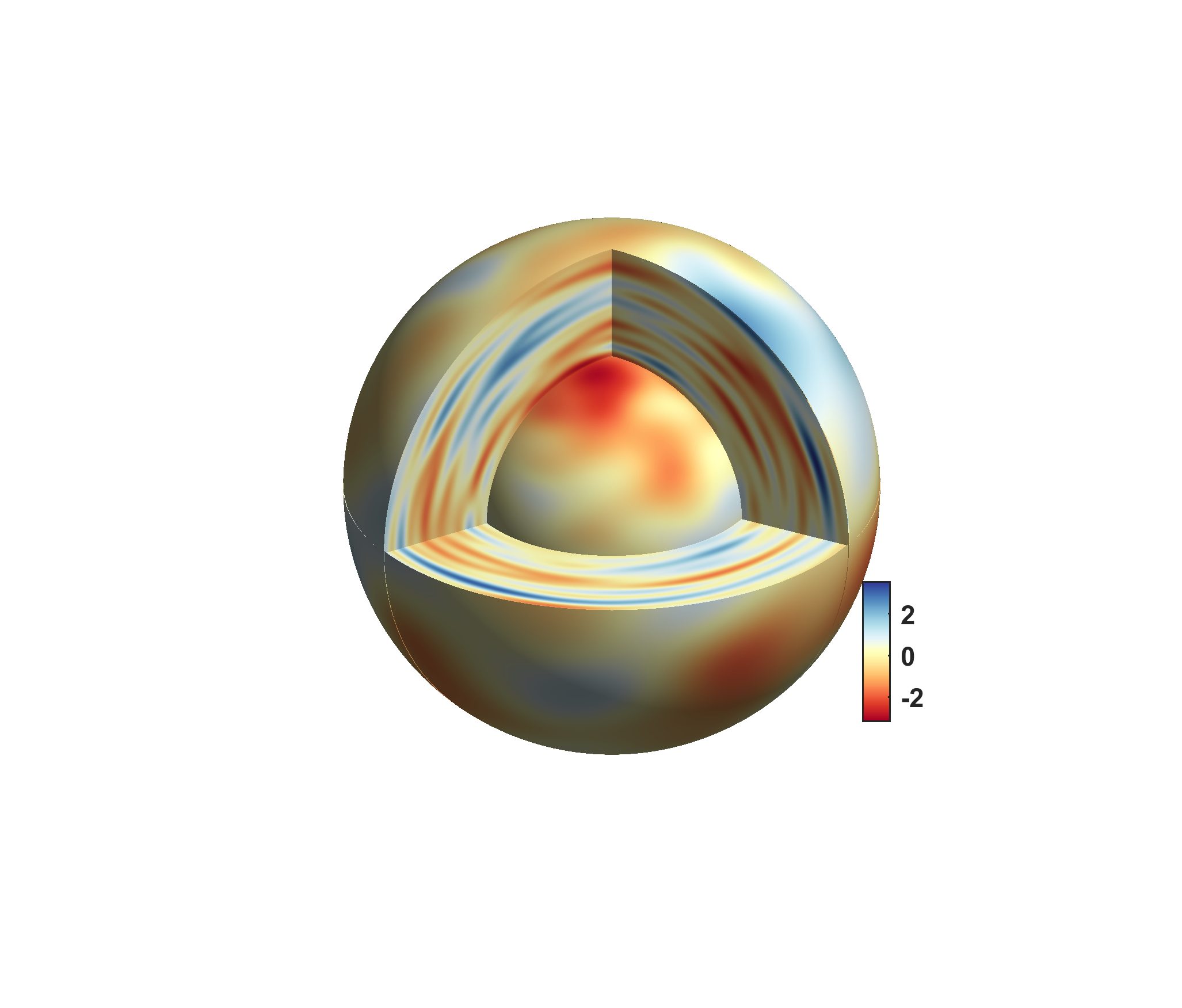}
		\caption{3D realizations of random field related for $I=0.05$ and $\rho=0.7$.}\label{fig:fig4}
	\end{center}
\end{figure}

\section{Conclusion}

We propose a new method for statistical modeling of the three-dimensional random field in a unit ball. This approach imposes strict conditions on the form of the covariance function (Eq. \ref{eqn:7}). It allows for a rigorous mathematical description of the simulated random field. The derived modeling formulas are tested by numerical computations. Examples of the generated random field realizations are presented.

\section*{Acknowledgments}
Dmitriy Kolyukhin gratefully acknowledges the financial support from RFBR and NSFB, project number 20-51-18009. Alexander Minakov acknowledges funding through the Research Council of Norway center of excellence funding scheme, project 223272 (The Centre for Earth Evolution and Dynamics). AM also acknowledges the project "3D Earth - A Dynamic Living Planet" funded by ESA as a Support to Science Element (STSE).
In this work, we used toolbox SHBUNDLE (v. 4/11.2018) that was developed by N. Sneeuw, M. Weigelt, B. Devaraju, M. Roth et. al. and provided via download from the Institute of Geodesy (GIS), University of Stuttgart:  https://www.gis.uni-stuttgart.de/en/research/downloads/shbundle/

\end{document}